\documentclass{article}
\usepackage{amsmath}
\usepackage{amssymb}
\usepackage{amsthm}
\newtheorem*{theorem}{Theorem}
\newtheorem*{corollary}{Corollary}
\newtheorem*{lemma}{Lemma}
\theoremstyle{definition}
\newtheorem{defn}{Definition}
\newcommand{\QTM}{Quantum Turing Machine}
\newcommand{\BQTM}{Bulk \QTM}

\newcommand{\qTM}{quantum Turing machine}
\newcommand{\bqTM}{bulk \qTM}
\newcommand{\mbqTM}{modified \bqTM}

\title{A Note on Bulk Quantum Turing Machine}
\author{MATSUI Tetsushi\footnote{Department of Mathematics Tokyo Metropolitan University}\\ \texttt{tetsushi@tnt.math.metro-u.ac.jp}}
\date{}
\begin{document}
\maketitle
\begin{abstract}
  Recently, among experiments for realization of quantum computers,
  NMR quantum computers have achieved the most impressive succession.
  There is a model of the NMR quantum computation, namely Atsumi and
  Nishino's \bqTM.  It assumes, however, an unnatural assumption with
  quantum mechanics.  We, then, define a more natural and quantum
  mechanically realizable \mbqTM, and show its computational ability
  by comparing complexity classes with \qTM's counter part.
\end{abstract}
\section{Introduction}
Recently, among experiments for realization of quantum computers, 
Nuclear Magnetic Resonance (NMR)
quantum computers have achieved the most impressive succession.  For
example, Vandersypen and his colleagues of IBM and Stanford University
compose organic molecules realizing $7$ quantum bits, and factor $15$ with
them using Shor's algorithm \cite{VS}.

The NMR quantum computation differs from the ordinary quantum
computation; since it manipulates many molecules simultaneously, one
cannot utilize the projection property of measurement but can only
obtain the ensemble average.  Note that, however, it is merely unusable
the projection in order to select a specific state, but in fact each
molecules is projected to the state when it is measured so that the
whole ensemble is the ensemble which gives the measurement value of the
ensemble average.  It is a basic assumption of the quantum mechanics.

Today, there is a model of the NMR quantum computation by Atsumi and
Nishino: \bqTM.
It has, however, two assumptions contradicting with quantum mechanics:
(1) a measurement does not cause projection on each molecules, and
(2) the probability to obtain the ensemble average value in a certain
interval is $1$.
We already mentioned about the first point above, and the second point
will be shown false later.

We shall define a new model of the NMR quantum computer removing the
difficulties of Atsumi and Nishino's model.
Then we shall show the computational power is the same with both Atsumi and
Nishino's {\bqTM} and the original {\qTM} for decision problems.

\section{Definitions} \label{secdef}
At first, we recall the definitions of the {\qTM} and {\bqTM}.
The first part of the definition is common to both.

\begin{defn}[Tape, Symbols and Unitarity] \label{basedef}
A {\qTM} (QTM) (or {\bqTM} (BQTM)) $\mathcal{M}$ is defined by a triplet $(\Sigma, Q, \delta)$
 where: $\Sigma$ is a finite alphabet with an identified blank symbol
 $\#$, $Q$ is a finite set of states with an identified initial state
 $q_0$ and final state $q_f$ which is not equal to $q_0$, and $\delta$
 is a function called quantum transition function:
\begin{equation}
 \delta : Q \times \Sigma \rightarrow \mathbf{\tilde{C}}^{\Sigma \times Q \times D}
\end{equation}
where $D$ denotes directions $\{L, R\}$ or $\{L,N,R\}$, and
 $\mathbf{\tilde{C}}$ denotes the set of computable numbers.
The {\qTM} (or {\bqTM}) has a two-way infinite
 tape of cells indexed by $\mathbf{Z}$ and a single tape head that moves
 along the tape.

The quantum transition function $\delta$ induces a time evolution
 operator $U_\mathcal{M}$ of the inner-product space of finite complex linear
 combination of configurations of $\mathcal{M}$.
The time evolution operator $U_\mathcal{M}$ must be unitary.
\end{defn}

The observation of {\QTM} is carried out as usual physics.
\begin{defn}[Observation of \QTM]
When a {\qTM} $\mathcal{M}$ in a superposition
$\Psi = \Sigma_i \alpha_i c_i$ is
observed, each configuration $c_i$ is obtained with probability
$|\alpha_i|^2$ and the superposition of $\mathcal{M}$ is updated to 
$\Psi' = c_i$.  A partial observation is also allowed.
\end{defn}

Atsumi and Nishino replaced the observation of {\qTM} above with the
measurement of {\bqTM} below in~\cite{AN}.

\begin{defn}[Measurement of \BQTM]
A measurement is an observation such that $|\alpha|^2 - |\beta|^2$ is
obtained for a qubit $\alpha |1\rangle + \beta |0\rangle$ with error
range less than $\theta$ and with probability 1.  The measurement does
not disturb the superposition which $\mathcal{M}$ is in, and can be
repeated several times.  No partial observation is allowed.
\end{defn}

We make a modification to the measurement.

\begin{defn}[($\epsilon$, $\theta$)-measurement] \label{etmes}
An ($\epsilon$, $\theta$)-measurement is an observation such that
$|\alpha|^2 - |\beta|^2$ is obtained for a qubit $\alpha |1\rangle +
\beta |0\rangle$ with error range less than $\theta$ and error probability
less than $\epsilon$.
If either $|\alpha|^2$ or $|\beta|^2$ is zero, the error probability is
 exceptionally zero.
The measurement does disturb the superposition which $\mathcal{M}$ is in, and
 can not be repeated multiple times.
No partial observation is allowed. 
\end{defn}

We call a quantum computer defined by definition~\ref{basedef} and
\ref{etmes} a {\mbqTM} (MBQTM).

Since a set of parallel statistically independent {\qTM}s gives an model
of \mbqTM; the bigger the number of {\qTM}s gets, the smaller the
parameters $\theta$ and $\epsilon$ of {\mbqTM} get. A quantitative
relationship among parameters is given in the following subsection, but
beforehand we give a qualitative statement.

\begin{lemma}
If a {\qTM} without partial measurements and resulting $0$ or $1$
exists, corresponding {\mbqTM} exists for $\frac{1}{2} > \forall \theta,
\epsilon > 0$.
\end{lemma}
\begin{proof}
Consider $n$ parallel independent {\qTM}s without partial measurements
and resulting $0$ or $1$.
Since partial measurements are not used, the computation is also carried
 out by {\mbqTM} having the same quantum transition function.
Therefore, the only difference is the final observation or measurement.
If the final superposition of the cell is
$\alpha |1\rangle + \beta |0\rangle$ and one assigns $-1$ to
 $|0\rangle$, the ensemble average is $|\alpha|^2 - |\beta|^2$.
The law of large numbers assures, the bigger $n$ gets, the smaller the
 error probability $\epsilon$ becomes with given value range of $\theta$.
\end{proof}

Note that the measurement of {\bqTM} is understood as the
(0,~$\theta$)-measurement, if we forget about the disturbance on the
superposition.
However, the parameters of positive $\theta$ with
$\epsilon = 0$ is not realized by the parallel independent {\qTM}s
until the superposition is in an eigen state ($|0\rangle$ or $|1\rangle$).

\subsection{Relationship among $\epsilon$, $\theta$ and the Number of
{\QTM}'s}

As stated above, {\mbqTM} is realizable by parallel independent
{\qTM}s and the parameters $\theta$ and $\epsilon$  of {\mbqTM} and the
number $n$ of {\qTM}s are dependent.
The relationship among them are known by de~Moivre - Laplace's theorem
asymptotically:
\begin{equation}
\frac{1}{\sqrt{2 \pi}} \int_{-t}^{t} e^{-\frac{x^2}{2}} dx \sim 1 - \epsilon
\end{equation}
where $t = 2 \theta \sqrt{n}$.

By this formula, the table \ref{tep} is obtained. The number $n$ is
considered as the number of molecules in NMR to be observed.

\begin{table}[tbp]
\caption{value of $n$ for $\theta$, $\epsilon$ \label{tep}}
\begin{center}
\begin{tabular}{|r|r|r|r|r|}
\hline
\multicolumn{2}{|c|}{value of} & \multicolumn{3}{c|}{$\epsilon$}\\
\cline{3-5}
\multicolumn{2}{|c|}{$n$} & $0.04550$ & $0.02000$ & $0.01000$\\
\hline
& $2^{-5}$ & $1024$ & $1699$ & $2018$ \\
\cline{2-5}
$\theta$ & $2^{-6}$ & $4096$ & $6795$ & $8069$ \\
\cline{2-5}
& $2^{-7}$ & $16384$ & $27177$ & $32275$ \\
\hline
\end{tabular}
\end{center}
\end{table}

\section{Complexity Classes}
There are some complexity classes for {\qTM} and corresponding classes
for {\bqTM}.  We shall define the corresponding classes for {\mbqTM},
then to show their equivalences.  Since we concern classes of decision
problems, we assume that the alphabet $\Sigma$ includes $\{0, 1\}$.
Moreover, a tape cell called acceptance cell be in the superposition
$\alpha|1\rangle + \beta|0\rangle$ when it is observed, i.e. there is no
possibility to have blank or any other symbols.

We shall discuss about three kinds of classes.
The classes for {\qTM} was defined by Bernstein and Vazirani~\cite{BV}, and
for {\bqTM} by Nishino et al.\ \cite{Ni}\cite{NSAS}.

\subsection{Exact Quantum Polynomial Time}

First of all, we see the classes of exact quantum polynomial time
languages.

\begin{defn}[$\mathbf{EQP}$, $\mathbf{EBQP}$, $\mathbf{EBQP^*}$]
The quantum complexity classes of exact quantum polynomial time
languages are defined according to the models of the quantum computers:
\begin{enumerate}
\item A language $L$ is in the class $\mathbf{EQP}$ if and only if there
      exists a {\qTM} and polynomial $p$ such that for any input $x$ an
      observation of a certain tape cell after calculation of $p(|x|)$
      steps gives $1$ with probability $1$ if $x$ belongs to $L$, $0$
      with probability $1$ otherwise.
\item A language $L$ is in the class $\mathbf{EBQP}$ if and only if
      there exists a {\bqTM} and polynomial $p$ such that for any input
      $x$ a measurement of a certain tape cell after calculation of
      $p(|x|)$ steps gives more than $1-\theta$ with probability $1$ if
      $x$ belongs to $L$, less than $-1+\theta$ with probability $1$
      otherwise.
\item A language $L$ is in the class $\mathbf{EBQP^*}$ if and only if
      there exists a {\mbqTM} and polynomial $p$ such that for any input
      $x$ an ($\epsilon$,$\theta$)-measurement of a certain tape cell
      after calculation of $p(|x|)$ steps gives more than $1-\theta$
      with probability $1$ if $x$ belongs to $L$, less than $-1+\theta$
      with probability $1$ otherwise.
\end{enumerate}
\end{defn}

\begin{theorem}
$\mathbf{EQP}$ and $\mathbf{EBQP^*}$ are equivalent.
\end{theorem}
\begin{proof}
It is possible to observe a value from a tape cell of {\qTM} with
probability $1$ only when the cell is equals to one of the eigen states;
$|0 \rangle$ or $|1 \rangle$.  Thus, if a language $L$ is in the class
$\mathbf{EQP}$, there exists a {\qTM} $\mathcal{M}$ with a certain cell
in the eigen state to be observed at the last step.  Suppose {\mbqTM}
$\mathcal{M^*}$ which has the same quantum transition function $\delta$
with $\mathcal{M}$.  The calculation steps are identical and the tape
cell to be read is in the eigen state.  Then, by the
definition~\ref{etmes}, it is possible to obtain the value with
probability $1$.  Thus, the language $L$ is in the class
$\mathbf{EBQP^*}$.

Conversely, if a language $L$ is in the class $\mathbf{EBQP^*}$, there
exists a {\mbqTM} $\mathcal{M^*}$.  Since the probability to observe
either one of the values $1$ or $-1$ is $1$, the tape cell is in either
one of the eigen state when it is observed.  Suppose {\qTM}
$\mathcal{M}$ which has the same quantum transition function $\delta^*$
with $\mathcal{M^*}$.  Then, the calculation steps are identical and the
final result is obtained with probability $1$, i.e. the language $L$ is
in the class $\mathbf{EQP}$.
\end{proof}

\begin{corollary}
$\mathbf{EBQP}$ and $\mathbf{EBQP^*}$ are equivalent.
\end{corollary}
\begin{proof}
It is a direct consequence of $\mathbf{EQP} = \mathbf{EBQP}$ \cite{Ni}.
\end{proof}

\subsection{Bounded Error Quantum Polynomial Time}

Next, we see the classes of bounded error quantum polynomial time
languages. These are the most important classes.

\begin{defn}[$\mathbf{BQP}$, $\mathbf{BBQP}$, $\mathbf{BBQP^*}$]
The quantum complexity classes of bounded error quantum polynomial time
languages are defined according to the models of the quantum computers:
\begin{enumerate}
\item A language $L$ is in the class $\mathbf{BQP}$ if and only if there
      exists a {\qTM} such that for any input $x$ an observation of a
      certain tape cell after calculation of polynomial time of its size
      gives $1$ with probability more than $2/3$ if $x$ belongs to $L$,
      $0$ with probability more than $2/3$ otherwise.
\item A language $L$ is in the class $\mathbf{BBQP}$ if and only if
      there exists a {\bqTM} such that for any input $x$ a measurement of
      a certain tape cell after calculation of polynomial time of its
      size gives more than $1/3-\theta$ if $x$ belongs to $L$, less than
      $-1/3+\theta$ otherwise.
\item A language $L$ is in the class $\mathbf{BBQP^*}$ if and only if
      there exists a {\mbqTM} such that for any input $x$ an
      ($\epsilon$,~$\theta$)-measurement of a certain tape cell after
      calculation of polynomial time of its size gives more than $1/3$
      if $x$ belongs to $L$, less than $-1/3$ otherwise.
\end{enumerate}
\end{defn}

\begin{theorem}
$\mathbf{BQP}$ and $\mathbf{BBQP^*}$ are equivalent.
\end{theorem}
\begin{proof}
Assume that $L$ is in the class $\mathbf{BQP}$.  By the definition there is a
{\qTM} $\mathcal{M}$ which accepts $L$.  We can assume that the tape
cell to be observed is in superposition
$\alpha|1\rangle + \beta|0\rangle$
and $p = |\alpha|^2 > 2/3$ if an input belongs to $L$.
By the lemma of section~ref{secdef},there is a {\mbqTM} $\mathcal{M^*}$
corresponding to $\mathcal{M}$ with $\theta < p - 2/3$.  Then, the
($\epsilon$,~$\theta$)-measurement of $\mathcal{M^*}$ gives a value in
 range $(|\alpha|^2 -|\beta|^2 -\theta, |\alpha|^2 -|\beta|^2 + \theta)$.
\begin{eqnarray}
|\alpha|^2 - |\beta|^2 - \theta &=& p - (1 - p) - \theta\\
&>& 2p - 1 - p + 2/3\\
&>& 1/3.
\end{eqnarray}

The other case is shown similarly.  Thus, we can conclude $L$ is in the
class $\mathbf{BBQP^*}$.

Conversely, Assume that $L$ belongs to $\mathbf{BBQP^*}$.  By the definition
there is a {\mbqTM} $\mathcal{M^*}$ which accepts $L$.  If an input
belongs to $L$, the ($\epsilon$,~$\theta$)-measurement of the acceptance
cell gives a value more than $1/3$.  With an identity $|\alpha|^2 +
|\beta|^2 = 1$, it gives $|\alpha|^2 > 2/3$.  In the case an input does
not belongs to $L$, it is shown in the same way that $|\beta|^2 > 2/3$.
Thus, considering a {\qTM} $\mathcal{M}$ which has the same quantum
transition function with $\mathcal{M^*}$ leads to the conclusion that $L$
is in the class $\mathbf{BQP}$
\end{proof}

\begin{corollary}
$\mathbf{BBQP}$ and $\mathbf{BBQP^*}$ are equivalent.
\end{corollary}
\begin{proof}
It is a direct consequence of $\mathbf{BQP} = \mathbf{BBQP}$ \cite{Ni}.
\end{proof}

\subsection{Zero Error Quantum Polynomial Time}

Finally, we see the classes of zero error quantum polynomial time
languages.

\begin{defn}[$\mathbf{ZQP}$, $\mathbf{ZBQP}$ and $\mathbf{ZBQP^*}$]
The quantum complexity classes of zero error quantum polynomial time
languages are defined according to the models of the quantum computers:
\begin{enumerate}
\item A language $L$ is in the class $\mathbf{ZQP}$ if and only if there
      exists a {\qTM} such that for any input $x$ an observation of a
      certain tape cell (halt cell) after calculation of polynomial time
      of its size gives $1$ with probability more than $1/2$ and then an
      observation of another cell (decision cell) gives $1$ with
      probability $1$ if $x$ belongs to $L$, $0$ with probability $1$
      otherwise.
\item A language $L$ is in the class $\mathbf{ZBQP}$ if and only if there
      exists a {\bqTM} such that for any input $x$ an observation of a
      certain tape cell (halt cell) after calculation of polynomial time
      of its size gives more than $0$ with probability $1$ and then
      either one of the following cases holds:
\begin{itemize}
\item measurement of another cell (accept cell) gives $1$ with
      probability $1$ if $x$ belongs to $L$,
\item measurement of one another cell (reject cell) gives $-1$ with
      probability $1$ if $x$ does not belong to $L$.
\end{itemize}

\item A language $L$ is in the class $\mathbf{ZBQP^*}$ if and only if
      there exists a {\mbqTM} such that for any input $x$ an observation
      of a certain tape cell (halt cell) after calculation of polynomial
      time of its size gives more than $0$ with probability more than
      $1-\epsilon$ and then either one of the following cases holds:
\begin{itemize}
\item ($\epsilon$,~$\theta$)-measurement of another cell (accept cell)
      gives $1$ with probability $1$ if $x$ belongs to $L$,
\item ($\epsilon$,~$\theta$)-measurement of one another cell (reject cell)
      gives $-1$ with probability $1$ if $x$ does not belong to $L$.
\end{itemize}
\end{enumerate}
\end{defn}

\begin{theorem}
$\mathbf{ZQP}$ and $\mathbf{ZBQP^*}$ are equivalent.
\end{theorem}
\begin{proof}
If a language $L$ is in the class $\mathbf{ZQP}$, there exists a {\qTM}
$\mathcal{M}$ and a polynomial $p$, which is a time estimation polynomial.
We construct a {\mbqTM} $\mathcal{M^*}$ from
$\mathcal{M}$ in the following way.

At first, we replace the
initialization steps of the decision cell of $\mathcal{M}$ with the
steps to initialize the accept cell to $|1\rangle$ and the reject cell
to $|0\rangle$. If there is no initialization steps in $\mathcal{M}$, we
insert the steps above.

The steps of $\mathcal{M^*}$ after initialization are identical to
$\mathcal{M}$ until it reaches to the step to write the result in
decision cell.

Finally, the writing step is replaced with those which write the same
result in both the accept cell and the reject cell: if $x$ belongs to
$L$ then the result is $|1\rangle$, otherwise $|0\rangle$.

The changes increase at most a constant $k$ steps.  Thus, after $p(|x|)
+ k$ steps ($\epsilon$,~$\theta$)-measurement of halt cell can give, if
one choose an appropriate $\theta$, more than $0$ with probability more
than $1-\epsilon$, since the corresponding observation of $\mathcal{M}$
gives $1$ with probability more than $1/2$. At the moment, if $x$
belongs to $L$, the accept cell has not been changed since the
initialization and reading $|1\rangle$ gives $1$ with probability $1$.
On the other hand, reject cell is not in the eigen state and it is
impossible to obtain $-1$ with probability $1$ by
($\epsilon$,~$\theta$)-measurement.  In the case when $x$
does not belong to $L$, the behaviors of the accept and reject cell are
switched and an ($\epsilon$,~$\theta$)-measurement of reject cell gives
$-1$ with probability $1$.  Thus, $L$ is in $\mathbf{ZBQP^*}$.

Conversely, we assume $L$ is in the class $\mathbf{ZBQP^*}$.  Then,
there exists a {\mbqTM} $\mathcal{M^*}$ to accept $L$.  Consider a
{\qTM} $\mathcal{M}$ which has the same quantum transition function with
$\mathcal{M^*}$, and identify the accept cell or the reject cell as the
decision cell.  Obviously, if an observation of the halt cell gives $1$,
the result is obtained correctly with probability $1$.  Moreover, the
probability that the observation of halt cell gives $1$ is more than
$1/2$, because $\mathcal{M^*}$ gives more than $0$ with probability more
than $1 - \epsilon$.  Therefore, $L$ is in the class $\mathbf{ZQP}$.
\end{proof}

\begin{corollary}
$\mathbf{ZBQP}$ and $\mathbf{ZBQP^*}$ are equivalent.
\end{corollary}
\begin{proof}
It is a direct consequence of $\mathbf{ZQP} = \mathbf{ZBQP}$ \cite{NSAS}.
\end{proof}

\section{Conclusion}
We construct a model of NMR quantum computation named {\mbqTM}.  It can
be realized as a set of statistically independent {\qTM}s and more
consistent with quantum physics than {\bqTM}, but still the
computational power is equivalent to that of {\qTM} and {\bqTM}.  Since,
the main difference between {\bqTM} and {\mbqTM} is the consistency with
quantum physics, it is better to replace {\bqTM} with {\mbqTM}.


\begin{thebibliography}{9}
\bibitem{AN} K. Atsumi and T. Nishino, ``Solving NP-complete problems
        and factoring problems by using NMR quantum computation'',
        Transaction of Information Processing Society Japan
        Vol.43 No. SIG 7(TOM 6) pp.10--18, September 2002.
\bibitem{BV} E. Bernstein and U. Vazirani, ``Quantum complexity theory'',
        SIAM J. Comput. Vol.26 No.5, pp.1411--1473, 1997.
\bibitem{DD} D. Deutsch, ``Quantum theory, the Church-Turing principle and
        the universal quantum computer'', Proc. R. Soc. Lond. A 400,
        pp.97--117, 1985.
\bibitem{Ni} T. Nishino, ``How to design efficient quantum
        algorithms'',
        Transaction of Information Processing Society Japan
        Vol.43 No. SIG 7(TOM 6) pp.1--9, September 2002.
\bibitem{NSAS} T. Nishino, H. Shibata, K. Atsumi, T. Shima, ``Solving
        function problems and NP-Complete Problems by NMR Quantum
        Computation'',
        Technical Report of IEICE COMP 98-71  pp.65--72, December 1998.
\bibitem{VS} L. M. K. Vandersypen, M. Steffen, G. Breyta, C. S. Yannoni,
        M. H. Sherwood, I. L. Chuang, Experimental Realization of
        Shor's Quantum Factoring Algorithm Using Nuclear Magnetic Resonance,
        Nature Vol.414 20/27 December pp.883--887, December 2001.
\bibitem{KZ} A. N. Kolmogorov, I. G. Zhurbenko, A. V. Prokhorov,
        Vvedenie v Teoriyu Veroiatnostei 2nd ed., Nauka, Moscow, 1995;
        T. Maruyama, Y. Baba (Japanese translation), Korumogorohu no
        kakuritsuronnyuumon, Morikita Shuppan, Tokyo, 2003.
\end{thebibliography}
\end{document}